# Birds, Frogs and the Measurement Problem

Stephen Boughn*
Departments of Physics and Astronomy, Haverford College, Haverford PA

**Preface**

Physicists and philosophers alike have pondered the measurement problem in quantum mechanics since the beginning of the theory nearly a century ago. Marcel Reginatto and I (2013) have made the case as have many others, Niels Bohr among them, that the problem is artificial and needs no resolution. The purpose of my present essay is not to elaborate on this conclusion but rather to shed light on the reasons why the problem arose in the first place and why it persists today, a more sociological than physical reflection. I am indebted to Freeman Dyson for his support of my forays into the foundations of quantum mechanics and, as a tribute to him, make use of his "birds and frogs" metaphor as explained below. I am not suggesting that he would have agreed with my ruminations nor approved of my use of his metaphor. On the other hand, from his writings and my conversations with him, it is clear that Dyson was as dismissive of the measurement problem as Bohr.

**The Measurement Problem**

Even though this is a thoroughly non-technical essay, it is incumbent upon me to provide at least a cursory account of the measurement problem. The statistical nature of quantum theory results in conflicts with a classical notion of reality and is in large part responsible for the quantum measurement problem. Here is, very roughly, how it comes about. Consider the simple case of a single particle. In the classical world, the state of the particle is expressed by its trajectory as a function of time, $x(t)$, a trajectory that satisfies Newton's laws. In the context of quantum mechanics, the state of the particle is given by a complex wave function of space and time, $\psi(x,t)$, that satisfies the Schrödinger equation. The success of Schrödinger's equation in describing the structure of the hydrogen atom led to its acceptance as a law of nature after which the pressing question was the physical significance of $\psi$. Following a suggestion by Einstein, Max Born postulated that the absolute value squared of the wave function, $|\psi(x,t)|^2$, corresponds to the probability density that the particle will be observed near the position $x$ at the time $t$. From then on, the probabilistic character of quantum mechanics was a fact of life and with it came the measurement problem.

---

* sboughn@haverford.edu

The simplest version is the following: If $|\psi(x,t)|^2$ is interpreted as the probability density for the particle being near the position $x$ at time $t$, then immediately after a measurement of its position at time $t_0$, the wave function of the particle must change to indicate that the particle is now located precisely at the measured position, $x_0$. This can be expressed as $\psi(x,t_0) \to \psi'(x,t_0)$ where $|\psi'(x,t_0)|^2 = \delta(x - x_0)$.[1] The problem is that Schrödinger's equation allows no such transition. What happened? How can the act of measuring the position of the particle somehow supervene a law of nature? In classical physics, the state $x(t)$ seems to present no such difficulty; however, we'll get back to this point later. There are many more subtle aspects of the measurement problem, but this simple example will suffice to frame the topic of my essay.

**Birds and Frogs**

In 2008, Freeman Dyson gave a lecture entitled *Birds and Frogs* to the American Mathematical Society in Vancouver. The following is the opening paragraph of that talk (Dyson 2015 p. 37):

"Some mathematicians are birds, others are frogs. Birds fly high in the air and survey broad vistas of mathematics out to the far horizon. They delight in concepts that unify our thinking and bring together diverse problems from different parts of the landscape. Frogs live in the mud below and see only the flowers that grow nearby. They delight in the detail of particular objects, and they solve problems one at a time. I happen to be a frog, but many of my best friends are birds. The main theme of my talk tonight is this. Mathematics needs both birds and frogs. Mathematics is rich and beautiful because birds give it broad visions and frogs give it intricate details. Mathematics is both great art and important science because it combines generality of concepts with depths of structures. It is stupid to claim that birds are better than frogs because they see farther, or that frogs are better than birds because they see deeper. The

---

[1] $\delta(x - x_0)$ is the Dirac delta function that fixes the particle's position at $x_0$.

world of mathematics is both broad and deep, and we need birds and frogs working together to explore it."

Dyson did not restrict his metaphor to mathematicians but also applied it to physicists, philosophers, and all other scientists for that matter. Among physicists that Dyson considered birds are Newton, Einstein, Schrödinger, Weyl, Ed Witten, and (in a separate talk) John Wheeler. Although he didn't mention Bohr and Heisenberg (remember this was a talk to mathematicians), I strongly suspect that he would have labeled them as birds. It's interesting that along with himself he characterized John von Neumann as a frog even though "everyone expected him to fly like a bird". (I'll come back to von Neumann later.) I'm an experimentalist and am certainly a frog as, I expect, are most experimenters with some notable exceptions. I imagine that Dyson would have labeled Ernst Mach, Enrico Fermi, and even one of my mentors, Robert Dicke, as birds. It's the case that birds occasionally descend down into the mud as is clearly evident with these three. So what does all this have to do with measurements? First, let me expand a bit more on the measurement problem.

**Back to the Measurement Problem**

The measurement induced transition referred to above, $\psi(x) \rightarrow \psi'(x)$ such that $|\psi'(x)|^2 = \delta(x - x_0)$, is often referred to as *wave function collapse* or *quantum state reduction*. Remember that this transition constitutes a violation of a law of physics, the Schrödinger equation. Heisenberg may have been the first to draw attention to this apparent phenomenon in his 1927 seminal paper on the indeterminacy relation, the Heisenberg Uncertainty Principle as we now call it. Dirac was more explicit in his 1930 influential text, *The Principles of Quantum Mechanics*, (Dirac 1958):

> When we measure a real dynamical variable $\xi$, the disturbance involved in the act of measurement causes a jump in the state of the dynamical system…if we make a second measurement of the same dynamical variable $\xi$ immediately after the first, the result of the second measurement must be the same as that of the first…This conclusion must still hold if the second measurement is not actually made. In this way we see that a measurement always causes the system to jump into an eigenstate of the dynamical variable that is being measured, the eigenvalue this eigenstate belongs to being equal to the result of the measurement.

It was von Neumann who first endeavored to give an entirely quantum mechanical description of a measurement in his 1932 equally influential text, *The Mathematical Foundations of Quantum Mechanics* (von Neumann 1955). He considered it manifest that the measuring apparatus was describable by quantum mechanics, i.e., by a quantum mechanical wave function. This led him to specify two processes in quantum mechanics: the discontinuous change of states that occurs as a result of a measurement (state reduction); and the unitary evolution of quantum states via the Schrödinger equation. He demonstrated that quantum mechanics was self-consistent including these two processes but didn't seem to be particularly bothered by the fact that only one of them was governed by a law of physics, that is, by the Schrödinger equation.

Bohr and Heisenberg also agreed that measuring apparatus could be described by quantum mechanics; however, unlike Dirac and von Neumann, neither of them found it necessary to invoke the process of wave function collapse. From a 2016 correspondence with Dyson, I know he was in firm agreement with Bohr on this matter[2]. So what led Dirac and von Neumann to the contrary point of view?

**Bird's-eye View of the Measurement Problem**

Einstein, Bohr, Schrödinger, and Heisenberg were high-flying birds who surveyed "broad vistas" of the quantum nature of matter and radiation "out to the far horizon". These physicists created[3] a quantum theory to describe what they saw. Dirac and von Neumann were frogs who set the stage for the measurement problem by characterizing wave function collapse as one of the two processes by which wave functions evolve, the other being the evolution described by the Schrödinger equation. (Although Dirac does not appear in Dyson's list of frogs, I suspect that he would have labeled Dirac as such.[4]) What did they, and others since then, see that led them to such a viewpoint? I suspect that it arises from a high altitude vantage point when these two frogs took to the air like birds. Looking down on the quantum landscape from up there, they see a quantum world of matter and radiation and a myriad of experimental frogs building instruments and making measurements. They also see many brilliant theorists, like Dyson, down in the mud solving problems in "intricate detail". Wave functions (quantum states) are everywhere. From on high, wave functions might be mistaken for the physical world they are only intended to describe. The trouble arises when one considers wave functions as part of the real physical world. Then when a measurement is made and the wave function either disappears or changes in a way not described by Schrödinger's equation, it seems imperative to investigate the details of this process.

---

[2] In response to an essay I wrote criticizing Bell's theorem, Dyson remarked (Dyson 2016), "I particularly like your last page, where you say that wave function collapse is not predicted by quantum mechanics. That is true, but I would like it better if you made a stronger statement. Wave function collapse never happens. What happens is that the wave function becomes irrelevant after an observation.

[3] I use the word "created" purposefully for quantum mechanics is, indeed, a human invention.

[4] Consider the following statement by Dirac (1963): "It may well be that the next advance in physics will come along these lines: people first discovering the equations and then needing a few years of development in order to find the physical idea behind these equations. My own belief is that this is a more likely line of progress than trying to guess at physical pictures." Sounds like a frog to me.

Wave functions are the tools of theorists in the same sense that laboratory instruments are the tools of experimentalists. It's certainly important to know and understand the types of tools one uses; however, there are questions about them that mustn't be asked. I'm an experimental physicist and my tools include voltmeters, oscilloscopes, lasers, etc. I certainly endeavor to understand and know how to use these tools. On the other hand certain questions about them make little sense. For example after having used a tool in the context of a given experiment, if I were asked where will the tool next appear in an experiment, I would answer: " I have no idea. I haven't even decided on what experiment to do next." Certainly no law of physics tells me what to do with my instruments. Wave functions are tools that theorists use to make predictions about the outcomes of experiments. Again certain questions about these tools make little sense. If a theorist were asked what is the wave function of any given quantum system she might also answer: "I have no idea. I haven't yet decided what quantum process I choose to investigate. After I do decide, I'll know how to prepare the quantum system and will then be able to specify the appropriate wave function." As before, no law of physics directs the theorist to decide what prediction she will eventually make.

I realize that I have tied the measurement problem to Dirac and von Neumann and then characterized them as frogs taking to the air like birds.[5] This may be unfair for there are undoubtedly birds who also take the measurement problem seriously; although, none immediately come to mind. Nevertheless, it's my impression that the majority of physicists, birds and frogs alike, seem to be not particularly interested in the measurement problem nor consider it to be an important unresolved dilemma.

---

[5] My intension is not to belittle von Neumann and Dirac. They were both brilliant theoretical physicists and mathematicians.

**Back to $x(t)$**

I began by drawing an analogy between the quantum mechanical wave function, $\psi(x,t)$, and the classical trajectory, $x(t)$, of a particle. Why has former caused such consternation while the latter has not? Certainly, the statistical aspect of $\psi(x,t)$ differentiates it from the precision implied by $x(t)$ and that difference provides part of the explanation. However, I think we're confusing two different aspects of the state of the particle. The function $x(t)$ can represent either a history of the past locations of the particle or a prediction of its future locations. On the other hand, $\psi(x,t)$ never represents a description of the past state of a particle but rather exclusively represents predictions of future observations. Dyson once remarked to me (Dyson 2016), "The wave function only describes probabilities for the result of an observation in the future. When the observation moves into the past, the result is described by facts instead of probabilities."

So if we wish to draw an analogy with quantum mechanics then we should stick to the interpretation of $x(t)$ as a prediction of future observations. Take for example a pool player predicting the trajectory of the 8 ball after it has been struck by the cue ball. The player presumably applies (intuitively) Newtonian mechanics to properly wield the cue stick. Like the analogous case with $\psi(x,t)$ in quantum mechanics, after the experiment, i.e., after the 8 ball falls into the corner pocket, the prediction represented by $x(t)$ becomes irrelevant. If one chooses to describe the trajectory of the 8 ball after the end of the experiment, $x(t)$ invariably represents the trajectory that is the past history of the ball and so has no predictive power. For example, this $x(t)$ can never be used to predict where the player places the 8 ball in the next rack, only the historical fact of where he did, in fact, place it. After all, the player may not yet have decided how to rack the balls or even if he or the other player will do the racking. So it seems that when used as a tool for prediction both $\psi(x,t)$ and $x(t)$ have limited domains. Even the statistical uncertainty inherent in $\psi(x,t)$ is not foreign to the predictive $x(t)$. As an experimentalist, I am well aware that no prediction, quantum or classical, is free from uncertainty which is one reason why "error bars" are ubiquitous in physics papers.

**Do Wave Functions Provide a Complete Description of Reality?**

I've just argued that the measurement problem can be dismissed primarily because wave functions have a limited domain of applicability. If this is the case, then it seems I am "forced to conclude that the quantum-mechanical description of physical reality given by wave functions is not complete". This was precisely the conclusion Einstein reached in the famous EPR paper (Einstein, Podolsky, & Rosen 1935). So be it. I'm quite willing to accept this charge as was Dyson (2016), "Einstein was right when he said that quantum mechanics gives an incomplete description of nature." In my last correspondence with him, Dyson (2019) wrote: "As you know, I disagree totally with the fashionable idea that quantum mechanics should be a complete description of nature. Following Niels Bohr, I consider the quantum world and the classical world to be equally real, two complementary views of nature."

**Final Remark**

Those of us who dismiss the measurement problem might be accused of ignoring an important dilemma that, if resolved, would facilitate a clearer understanding of nature. However, I suspect that Dyson would resist the notion of tying up our understanding of the world in a neat little package. In the same *Birds and Frogs* address to the American Mathematical Society that I quoted above, Dyson referred to Kurt Gödel's proofs that devastated Hilbert's program to build a unique and solid foundation for mathematics. He clearly reveled in Gödel's achievement, as did von Neumann (Dyson 2015, p. 49): "After Gödel, mathematics was no longer a single structure tied together with a unique concept of truth, but an archipelago of structures with diverse sets of axioms and diverse notions of truth. Gödel showed that mathematics is inexhaustible." As evidence for Dyson's analogous sentiment about physics, I leave you with words from his contribution to a 2002 symposium in honor of John Wheeler, (Dyson 2015, p. 320):

> The structure of theoretical physics as a whole, and of quantum theory in particular, looks to me like a makeshift agglomeration of bits and pieces, not like a finished design. If the structure of science is still provisional, still growing and changing as the years go by, it makes no sense to impose on the structure a spurious philosophical coherence. That is why I am skeptical of all attempts to squeeze quantum theory into a clean and tidy philosophical doctrine. I prefer to leave you with the feeling that we still have a lot to learn.

## Acknowledgements

I thank my two muses, Freeman Dyson and Marcel Reginatto, for encouraging my pursuit of understanding the foundations of quantum mechanics.